%% file: short.tex
\documentclass[conference]{IEEEtran}
\IEEEoverridecommandlockouts
\usepackage{cite}
\usepackage{amsmath,amssymb,amsfonts}
\usepackage{algorithmic}
\usepackage{graphicx}
\usepackage{textcomp}
\usepackage{xcolor}
\usepackage{xurl}
\usepackage{hyperref}
\def\BibTeX{{\rm B\kern-.05em{\sc i\kern-.025em b}\kern-.08em
    T\kern-.1667em\lower.7ex\hbox{E}\kern-.125emX}}
\begin{document}

\title{Association between Player Perception and AI Disclosure on Steam: An Exploratory Study
}

\author{
\IEEEauthorblockN{Fabian Wiesinger}
\IEEEauthorblockA{
\textit{Graz University of Technology} \\
fabian.wiesinger@student.tugraz.at
}
\and
\IEEEauthorblockN{Johanna Pirker}
\IEEEauthorblockA{
\textit{Technical University of Munich}\\
\textit{Graz University of Technology} \\
johanna.pirker@tugraz.at
}
\and
\IEEEauthorblockN{Denis Helic}
\IEEEauthorblockA{
\textit{Graz University of Technology} \\
dhelic@tugraz.at
}
}

\IEEEoverridecommandlockouts

\IEEEpubid{\makebox[\columnwidth]{979-8-3315-9476-3/26/\$31.00 \copyright2026 IEEE\hfill}
\hspace{\columnsep}\makebox[\columnwidth]{ }}

\maketitle

\IEEEpubidadjcol

\begin{abstract}
The use of generative AI in game development is growing rapidly and is the subject of intense debate in which player reactions play a central role. This paper contributes to this discussion by analyzing Steam reviews to examine whether games with AI disclosure are associated with different review patterns. Using a dataset across multiple disclosure groups, we apply linguistic analysis including keyword-based, word frequency, and characteristic word analysis to investigate the AI-related discussion on Steam. Our results show that AI-disclosed games are associated with higher levels of AI-related criticism in negative reviews. Our findings provide a first exploratory step towards understanding how players perceive AI disclosure.

\end{abstract}


\section{Introduction}

Artificial intelligence (AI) systems already produce text, images, and other content that are used in creative industries such as art, music, or digital media \cite{b11,b12,b13}. This development is particularly effective in the gaming industry for creating graphics, writing dialog, or designing game environments.
Currently, we find various examples of AI usage in digital games including programming, concept art, dialogue drafts, or localization support \cite{b11,b12,b13} indicating that AI-assisted workflows are increasingly becoming a part of development practices. At the same time, the growing use of generative AI in games has also sparked negative discussions around originality, artistic value, or ethical concerns such as copyright or the replacement of human creative labor.

In response to this development, Valve Corporation introduced a mandatory policy on AI disclosure on Steam in January 2024. According to this policy, developers must indicate whether generative AI methods were used during development and provide a description of their use, distinguishing between pre-generated and live-generated content.\footnote{\href{https://steamcommunity.com/groups/steamworks/announcements/detail/3862463747997849619}{Valve: AI Content on Steam}}\textsuperscript{,}\footnote{\href{https://partner.steamgames.com/doc/gettingstarted/contentsurvey}{Valve: Content Survey}}
Although transparency supports informed decisions, public discussion of this policy has also raised concerns that AI labels could be associated with stronger negative reactions from players.\footnote{\href{https://www.pcgamer.com/software/ai/epic-boss-tim-sweeney-thinks-stores-like-steam-should-stop-labelling-games-as-being-made-with-ai-it-makes-no-sense-he-says-because-ai-will-be-involved-in-nearly-all-future-production/}{PCGamer: Epic Boss Tim Sweeney on Steam AI Disclosure Policy}} On a general level, this debate can be linked to broader research on algorithmic aversion, suggesting that people may evaluate algorithmically supported results more critically once the algorithmic usage is disclosed \cite{b3}.

However, despite the growing importance of AI in game development, we still observe a gap in empirical research on players' reaction to AI-generated content on gaming platforms. In particular, there are currently no research results on the relation between Steam's disclosure notices and whether these notices are associated with, among others, more explicit AI-related criticism in user reviews. With our study, we fill this gap by conducting an exploratory analysis of a large dataset of Steam reviews from games with and without AI disclosure.
More specifically, our goal is to answer the following research question: \textit{Do games with AI disclosure receive more AI-related criticism than comparable games without AI disclosure?}

We answer our research question with a large-scale Steam review analysis of games with and without AI disclosure.
Specifically, we combine linguistic analysis including keyword-based analysis or analysis of characteristic words with a range of statistical and robustness tests to describe how review discourse differs between groups.
With our approach, we concentrate on associations between rating language and AI disclosure and not on causal effects. As such our study represents an initial step in characterizing the relation between player perception and AI disclosure on Steam.

We find that games with AI disclosure are associated with higher levels of AI-related criticism in negative Steam reviews than comparable games without AI disclosure. This pattern is robust across review-threshold settings and  resampling strategies. With our work, we make two contributions. First, our results can serve as a scientific basis for the construction of hypotheses for stronger future causal studies. Second, we make a practical contribution towards platform operators and game developers that may want to mitigate the increased criticism of games with AI content by, for instance, further increasing transparency about the actual content and the amount of AI-generated content. In addition, we publish all of our code.\footnote{\url{https://github.com/Fabisitooo/ba-thesis-steam-reviews}}

\section{Dataset}

\begin{table}[b]
\caption{Basic Dataset Statistics}
\begin{center}
\begin{tabular}{|c|c|c|c|c|}
\hline
\textbf{Group} & \multicolumn{2}{|c|}{\textbf{\#Games with reviews}} & \multicolumn{2}{|c|}{\textbf{\#Reviews}}\\
\cline{2-5}
 & \textbf{\textit{Negative}} & \textbf{\textit{Positive}}  & \textbf{\textit{Negative}} & \textbf{\textit{Positive}} \\
\hline
AI-labeled & 70 & 70  & 5,665 & 5,456 \\
\hline
Without AI label & 19 & 9 & 936 & 697 \\
\hline
Total & 89 & 79  & 6,601 & 6,153 \\
\hline
\end{tabular}
\label{tab_dataset_games}
\end{center}
\end{table}

We first identify games with AI disclosure labels, which we call \textit{AI-labeled} games from the Steam store page and platform metadata. We then retrieve all reviews for those identified games by Web scraping and
public Steam rating endpoints.
To make the groups with and without AI labels better comparable to each other, we then collect a set of games \textit{Without AI label} group based on Steam categories (tags) and popularity criteria. Particularly, we select ten most common categories in the AI-disclosed set including genres (Simulation, Casual, Adventure, Strategy, Action, RPG, and Sports) and publisher (Indie, Free To Play, and Early Access), and then query non-AI candidates within these categories retaining only games with at least 30 (to reduce noise) and at most 5000 (to reduce the popularity bias) reviews. Although we do not conduct a one-to-one matching where each AI game is paired with exactly one control game, with this design choice we aim at reducing distributional differences between AI-disclosed and non-disclosed titles at least with respect to game category and game popularity. However, we recognize that no suitable design can eliminate all potential confounding factors.

\begin{figure}[!t]
\centerline{\includegraphics[width=\columnwidth]{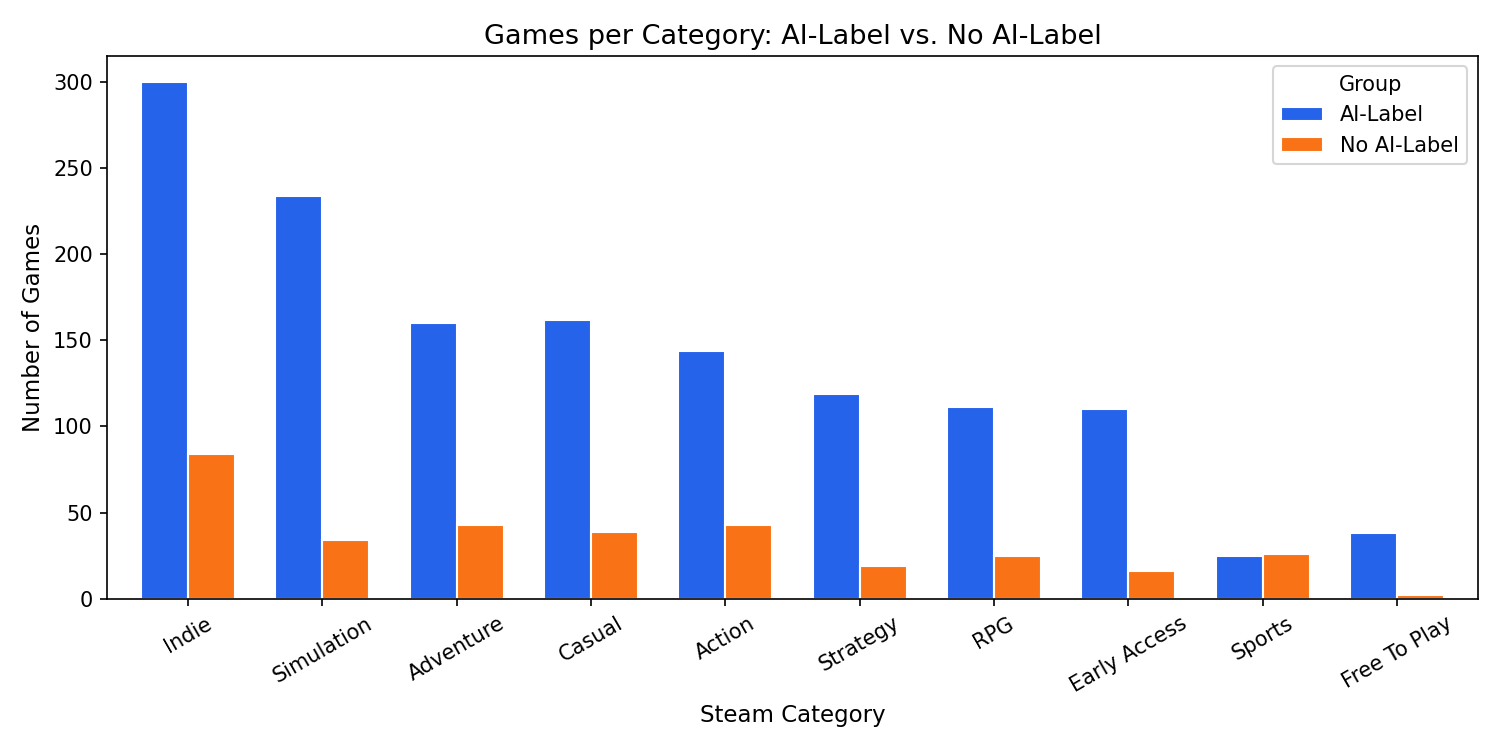}}
\caption{Distribution of games over Steam categories in our dataset. We show games with AI label in blue and games without AI label in orange.}
\label{categories}
\end{figure}

\begin{figure}[!t]
\centerline{\includegraphics[width=\columnwidth]{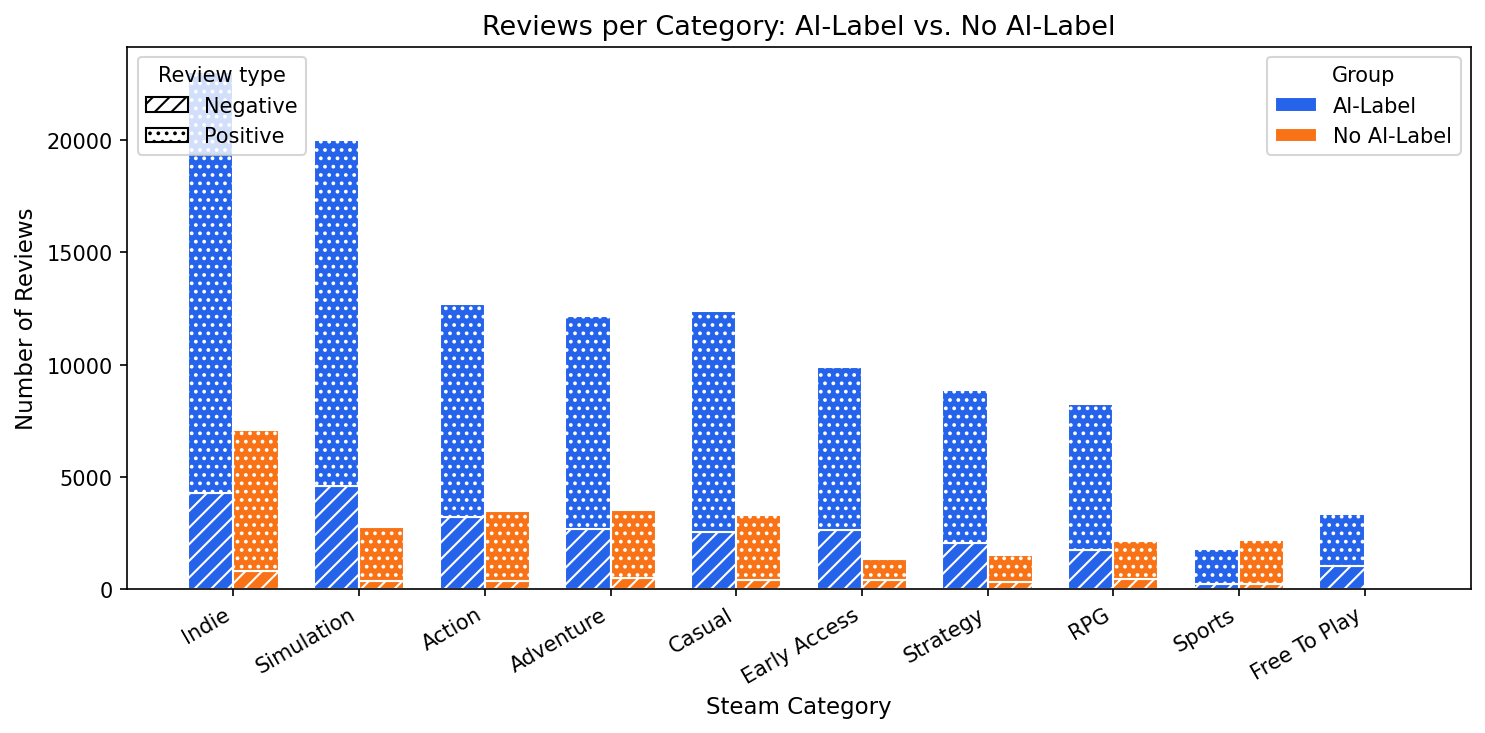}}
\caption{Distribution of game reviews over Steam categories in our dataset. We show positive and negative reviews with different hatch styles.}
\label{reviews}
\end{figure}

Our collected dataset comprises a total of 50,153 reviews (10,187 negative and 39,966 positive) for 602 games. We show some of the descriptive statistics of our datasets in Figures \ref{categories} and \ref{reviews}. Lastly, as we concentrate on the analysis of AI-related criticism we further filter the games and select only games with at least $30$ negative reviews. Our final dataset contains 6,601 negative reviews and 6,153 positive reviews including 5,665 negative and 5,456 positive reviews for \textit{AI-labeled} games and 936 negative and 697 positive reviews for games \textit{Without AI label}.
Table~\ref{tab_dataset_games}
summarizes the dataset that we use in this study.

\section{Methods}

\noindent\textbf{AI-Related Keyword Counts.}
First we check for the existence of AI-related criticism in negative reviews by counting AI-related keywords containing general terms from three areas: technical AI references such as ``AI'', ``generative'', ``ChatGPT'', ethics and copyright such as ``stolen'', ``plagiarism'', or ``copyright'', and quality-related criticism often associated with AI debates such as ``sloppy'', ``soulless'', or ``lazy'', as well as narrower AI-specific keywords such as ``AI-generated'', ``generated by AI'', ``midjourney'', or ``stable diffusion''. We construct this keyword set iteratively from
recurring terms from public debates around generative AI in creative work.

\noindent\textbf{Word Frequency Analysis.}
To compare words from negative and positive reviews, we analyze their frequencies in these reviews. We start by cleaning the review texts by removing standard English stop-words such as the'', ``and'', or ``is'', as well as high-frequency gaming-related terms such as ``game'', ``play'', ``hours'', or ``really''. We proceed by tokenizing the words using a custom regex tokenizer that keeps lowercase alphabetic and alphanumeric tokens, strips apostrophes, and enforces a minimum token length of three characters. We then compute word frequencies separately for positive and negative reviews and overall frequencies.

\noindent\textbf{Characteristic Words.}
Next, we analyze the characteristic language used in negative vs. positive reviews \cite{koncar2020, koncar2022}. In particular, we identify words that are substantially and significantly differently used in negative vs. positive reviews. Specifically, for each word that appears in at least $20$ reviews, we first construct a $2 \times 2$ contingency table with the number of reviews in which a given word appears as well as the number of reviews in which this word does not appear for both negative and positive reviews. By computing the chi-square significance tests and the odds ratios for all words, this contingency table allows us to compare whether the usage of a given word is significantly different in negative or positive reviews.
As we compute the significance test for multiple words simultaneously, we adjust the p-values using the Benjamini-Hochberg method to correct for the false discovery rate \cite{b16}.

\noindent\textbf{Linguistic Analysis with LIWC.}
We additionally use LIWC-22 \cite{tausczik2010} to compare linguistic category scores between reviews of games with and without AI disclosure. For each review, LIWC provides standardized scores for categories such as \textit{Authentic}, \textit{Tone}, words per sentence (WPS), \textit{BigWords} (words with more than six characters), cognitive processes, affect, and positive and negative emotion. We compute group means and standard deviations for each selected LIWC variable.

\noindent\textbf{Robustness Checks.}
In our study, we use a threshold on the minimum number of negative reviews per game. We set this threshold to $30$ but make additional computations by slightly varying this threshold to $10$, $20$, $40$, and $50$. We do not observe any substantial qualitative or quantitative changes with any of the alternative threshold settings.
Lastly, we quantify the uncertainty of our estimates by cluster bootstrap sampling at the game level \cite{efron1994}. Specifically, we resample the games 10,000 times with replacement within each group and use these resamples to calculate the bootstrap confidence intervals.

\section{Results}

\input{liwc_results_table}

\noindent\textbf{AI-Related Keyword Counts.}
AI-related keywords are more frequent in negative reviews of \textit{AI-labeled} games with the rate 22.26\% and bootstrap confidence intervals [18.35, 26.62] than in games \textit{Without AI label} with the rate 17.20\% [10.27, 24.50], suggesting the presence of AI-related criticism.
The difference between the rates of AI-related keywords between game groups is +5.07 percentage points [-2.97, 12.90].
Because the confidence interval for the collapsed two-group difference contains zero, we interpret this result as a positive descriptive association rather than as a statistically significant group difference under game-level bootstrap sampling.
Lastly, the proportion of negative reviews containing at least one AI-related keyword is 24.31\% in the \textit{AI-labeled} group and 14.35\% in the group \textit{Without AI label}. This finding supports the interpretation that our results are not limited to a single operationalization of AI-related criticism.

\begin{figure}[!t]
\centerline{\includegraphics[width=\columnwidth]{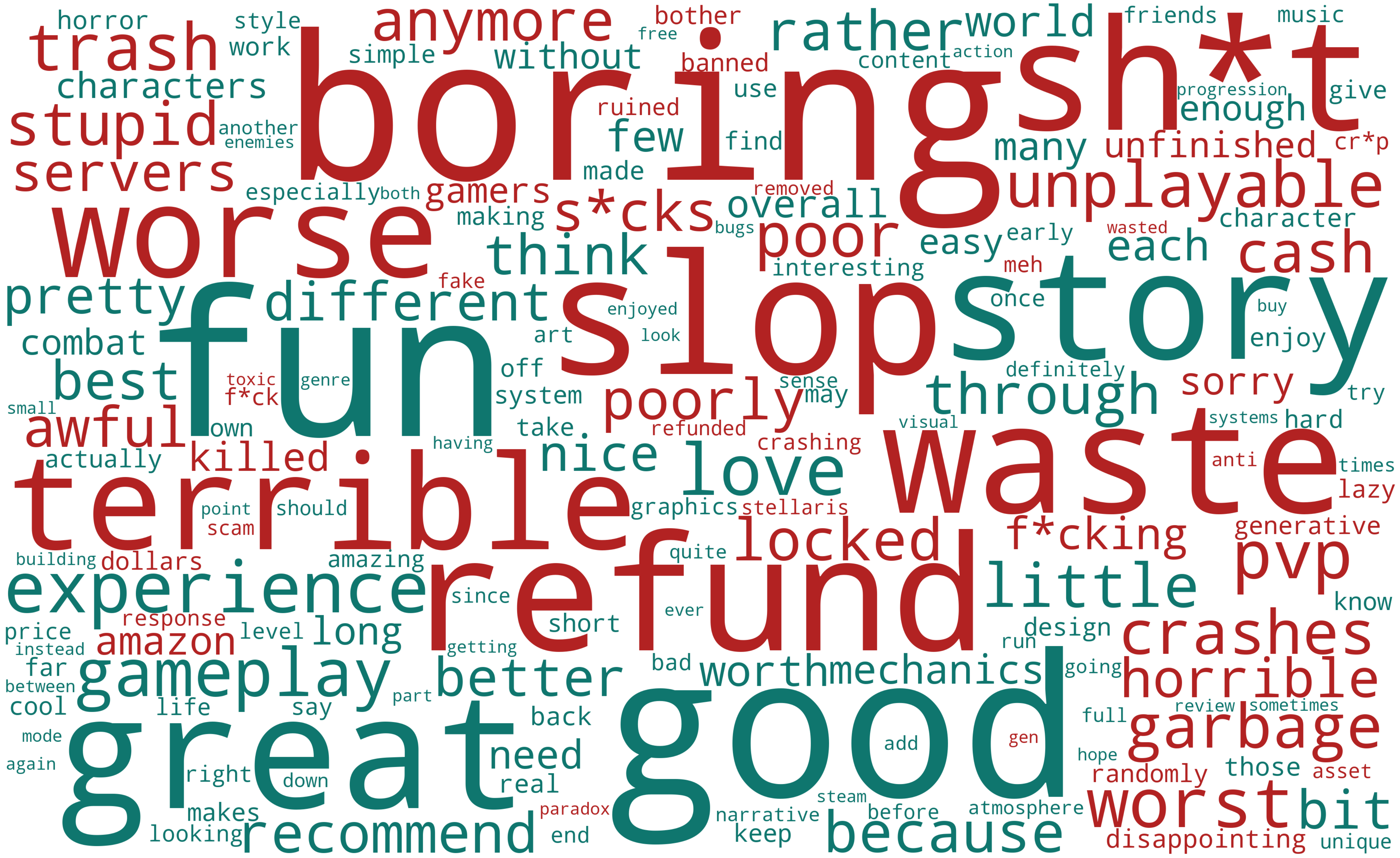}}
\caption{Combined overall word cloud for negative and positive reviews. Negative words are shown in red and positive words in teal. Word size reflects frequency normalized within each review category.}
\label{fig_wordcloud_combined}
\end{figure}

\noindent\textbf{Word Frequencies.}
The word frequency analysis also shows a clear contrast between the negative and positive review vocabulary. In negative reviews, frequently used words include ``boring,'' ``slop,'' ``refund,'' ``waste,'' and ``terrible.'' while in positive reviews, words such as ``fun,'' ``good,'' ``great,'' ``story,'' and ``love'' occur frequently.
Figure~\ref{fig_wordcloud_combined} visualizes the combined overall word cloud, where the most frequent negative terms are shown in red and the most frequent positive ones in teal.

\noindent\textbf{Characteristic Words.}
The contingency table analysis also shows that the vocabulary used in negative and positive reviews differs substantially from each other. Altogether, 7,697 words met the threshold for document frequency, and 51.19\% of these differ significantly in their usage between positive and negative reviews, after correcting for the false discovery rate. The significant proportion of words is 51.52\% for \textit{AI-labeled} games
and 37.73\% for games \textit{Without AI label}.

Overall, the words used more often in negative reviews include ``sh*t'' (odds ratio 2420.987), ``f*cking'' (900.531), or ``f*ck'' (772.357), while the words used more frequently in positive reviews include ``playful'' (0.011), ``enhances'' (0.012), or ``embraces'' (0.014).
Regarding words that are potentially related to AI, in the \textit{AI-labeled} group we observe terms such as ``generative'' (odds ratio 13.889), ``soulless'' (13.192), ``slop'' (10.977), ``lazy'' (6.491), and ``stolen'' (3.494), which are used more often in negative than in positive reviews. In the \textit{Without AI label} group, the strongest comparable negative-leaning examples are more general quality terms such as ``trash'' (12.767), ``garbage'' (7.921), or ``cheap'' (3.816), while direct AI-related terms are less prominent.
While among the top significant words we do not observe words directly related to the AI topic, the language used in the game reviews for games from both groups differs substantially from each other.

\noindent\textbf{Results with LIWC.}
The LIWC-22 analysis provides an additional linguistic comparison between reviews of games with and without AI disclosure. Table~\ref{tab:liwc_results} reports group means, standard deviations, and bootstrap confidence intervals for the difference between the AI-labeled and non-labeled group for a selection of LIWC variables. The strongest difference appears for \textit{Tone}, which is lower in the AI-labeled group, indicating a more negative overall emotional tone. In addition, affect-related categories such as \textit{Affect}, \textit{tone\_neg}, \textit{emo\_pos}, and \textit{emo\_neg} are higher for the AI-labeled group, while \textit{Authentic}, \textit{WPS}, and \textit{BigWords} show no substantial difference.

\noindent\textbf{Sensitivity Analysis.}
We observe stable results across thresholds. The AI-related keyword rates for \textit{AI-labeled} games range from 20.99\% (threshold 10) to 23.77\% (50), while those for games \textit{Without AI label} from 14.78\% (10) to 21.53\% (50).

\section{Discussion}

The review texts show that some players explicitly mention the use of AI. Terms such as ``AI'', ``AI-generated'', and ``generative'' appear directly in negative reviews, especially in the \textit{AI-labeled} group. This suggests that AI disclosure already became a part of the language users use to rate and discuss games.
The criticism in the game reviews does not appear to be purely technical in nature. In addition to direct references to AI, the language of the reviews frequently includes quality-related and normative terms such as ``slop,'' ``trash,'' ``garbage,'' and ``lazy.'' In some cases, ethical terminology also comes up, for example, in relation to copyright, plagiarism, or theft. This pattern is consistent with the broader public debate on authenticity, originality, and artistic work in AI-assisted creative production \cite{b14}. However, we do not claim that all cases of these terms refer exclusively to AI. The interpretation remains descriptive and is based on identified patterns.

For game developers, these findings suggest that AI disclosure can become part of the product evaluation itself rather than remain a neutral production detail. Once AI use is visible on the store page, players may interpret it through existing debates about quality, authenticity, and creative labor \cite{b14}. This means that developers using AI-assisted workflows may need to manage not only the technical integration of AI, but also the communication and transparency around AI usage.

\noindent\textbf{Limitations.}
Our study has several limitations. First, the keyword-based analysis cannot fully capture the linguistic context and false positives are still possible. Second, Steam reviews are selective and therefore do not equally reflect the opinions of all players.  Third, the data set is limited to the Steam platform and the results may not generalize to other gaming platforms.
Finally, this is a comparative and descriptive study. We do not control for confounders such as game quality or user prior exposure to AI labels. Hence, we only identify correlations between disclosure and the discourse in game reviews but do not establish any causal links.

\noindent\textbf{Further related work.}
Studies on public reactions to generative AI repeatedly reveal concerns about originality, effort, authenticity, and property rights. In the visual arts, literature, and music, public debate often revolves around whether AI-assisted works are less human, less creative, or ethically problematic due to copyright and training data issues \cite{b3,b14}.
In gaming, the general player perception of algorithmically generated content in digital games has been already investigated with player reviews. Previous studies have shown that review data can capture social influence, public discourse, review dynamics, and coordinated negativity in digital review environments \cite{b4,b5,b6,b7,b8,b21}. In our study, we build on this line of research and analyze the relationship of AI disclosure policy on Steam and the linguistic properties of player reviews.

\section{Conclusion}

This study analyzed a large dataset of Steam reviews to investigate the relationship between games with AI disclosure and the language used in players' reviews. Across various comparison strategies, the \textit{AI-labeled} group shows the strongest indication of AI-related criticism, while the \textit{Without AI label} group has the lowest score.
Frequency analyses, AI-driven keyword filtering, contingency tables, LIWC analysis and bootstrap uncertainty tests all support the conclusion that AI-disclosed and non-disclosed games are associated with measurably different patterns in review discourse. The study thus provides a descriptive and comparative analysis of how the disclosure of AI is reflected in player language on Steam without claim on causal interpretations.

\noindent\textbf{Future Work.}
Several expansions are possible in future work. Topic modeling could help identify recurring thematic structure beyond manually selected keywords. More semantically rich models based on transformer architectures could also support review-level or sentence-level classification of AI-related stances.
Longitudinal studies could examine how reactions develop after AI information has been disclosed on a store page or after updates have changed AI usage in a game. Finally, a cross-platform analysis would clarify whether the associations observed here are specific to Steam or reflect a broader pattern within gaming communities.

\bibliographystyle{IEEEtran}
\bibliography{references}

\end{document}

%% file: liwc_results_table.tex

\begin{table}[b]
\caption{LIWC-22 Analysis}
\label{tab:liwc_results}
\centering
\resizebox{\columnwidth}{!}{%
\begin{tabular}{|l|c|c|c|c|}
\hline
Metric & AI-Lab. & W/O-Lab. & Diff. & 95\% CI \\
\hline
Authentic & 50.06 (34.17) & 50.34 (34.66) & -0.27 & [-1.11, 0.57] \\
Tone & 69.31 (36.14) & 75.26 (33.25) & -5.95 & [-6.76, -5.14] \\
WPS & 13.85 (12.24) & 13.69 (11.24) & 0.16 & [-0.10, 0.40] \\
BigWords & 16.81 (14.40) & 16.95 (13.40) & -0.14 & [-0.43, 0.17]  \\
Cognition & 20.75 (12.70) & 20.51 (12.01) & 0.24 & [-0.03, 0.50] \\
Affect & 12.51 (16.48) & 11.88 (13.71) & 0.63 & [0.32, 0.95] \\
tone\_pos & 9.92 (15.78) & 9.66 (13.02) & 0.26 & [-0.05, 0.55] \\
tone\_neg & 2.18 (5.89) & 1.80 (4.56) & 0.38 & [0.27, 0.48] \\
emo\_pos & 5.19 (13.48) & 4.48 (10.56) & 0.71 & [0.46, 0.96] \\
emo\_neg & 1.03 (4.56) & 0.77 (3.60) & 0.26 & [0.17, 0.34] \\
\hline
\end{tabular}%
}
\end{table}